\documentclass[aps,pra,preprint,groupedaddress,showpacs]{revtex4-1}
\usepackage{amssymb}
\usepackage{amsmath}
\usepackage{dcolumn}
\usepackage{bm}
\usepackage{graphicx}
\usepackage{mathrsfs}

\usepackage{color}
\usepackage{ulem}
\usepackage{multirow}
\usepackage{multirow}
\usepackage{threeparttable}
\usepackage{subfigure}

\begin{document}

\title{Demonstration of quantum permutation algorithm with a single photon ququart}
\author{Feiran Wang}
\author{Yunlong Wang}
\author{Ruifeng Liu}
\author{Dongxu Chen}
\author{Pei Zhang}
\email{zhangpei@mail.ustc.edu.cn}
\author{Hong Gao}
\author{Fuli Li}
\address{MOE Key Laboratory for Nonequilibrium Synthesis and Modulation of Condensed Matter, and Department of Applied Physics, Xi'an Jiaotong University, Xi'an 710049, China.}
\begin{abstract}
We report an experiment to demonstrate a quantum permutation determining algorithm with linear optical system. By employing photon polarization and spatial modes, we realize the quantum ququart states and all the essential permutation transformations. Compared with the classical case, this work determines the parity of the permutation in only one step of evaluation and displays the remarkable speedup of quantum algorithm. This experiment is accomplished in single photon level and exhibits strong universality in quantum computation.

\bigskip
\bigskip

\end{abstract}

\maketitle

\address{MOE Key Laboratory for Nonequilibrium Synthesis and
Modulation of Condensed Matter, and Department of Applied Physics, Xi'an
Jiaotong University, Xi'an 710049, People's Republic of China}

\noindent \textbf{Introduction}

As quantum counterpart of classical computer, quantum computer reveals incredible efficiency to execute arithmetic tasks and threatens the security of classical communication.  Quantum algorithm is the sole of quantum computation, and it shows the amazing power of quantum parallelism and utilizes quantum interference to get the expected result, therefore it attracts particular concern to develop new quantum algorithm in recent years. The concept of simulating physics progresses with quantum computers was originated in Richard Feynman's observation that computers built from quantum mechanical components would be ideally suited to simulating quantum mechanics \cite{Feynman1982}. Since then, the first efficient quantum algorithm was proposed by Deutsch in 1985 \cite{Deutsch1985} and generalized by Deutsch and Jozsa in 1987 \cite{Deutsch1987}. Lately, an increasing number of practical programs are presented, such as factoring large integer \cite{Shor1994}, Grover's searching algorithm for database \cite{Grover1996} and Simon's exponential acceleration algorithm for the black box problem \cite{Simon1994}. What's more, Harrow \textit{et al.} in Bristol came up with a quantum scheme to decrease the computational complexity of solving linear system of equations from $ O (n) $ to $ log (n) $, and this is the first quantum algorithm to work out the most fundamental problems in engineering science \cite{Harrow2009}. Some  quantum algorithms have been demonstrated in different physical systems, such as ion traps \cite{Cirac1995,Blatt2008,Lanyon2011,Lanyon2013}, superconducting devices \cite{Nakamura1999,DiCarlo2009,Bialczak2010}, optical lattices \cite{Jaksch2004,Bloch2008}, quantum dots \cite{Fushman2008,Berezovsky2008}, and linear optics \cite{Knill2001,O'Brien2003,Mohseni2003,Zhang2010,Tame2014}.  Linear optical system is potential to realize quantum computer, due to its good scalability, easy-handling and high stability, and  it is a good candidate for implementing quantum algorithms.

 A new algorithm based on quantum Fourier transformation has been proposed to identify the parity of a permutation recently \cite{Gedik2014}. The permutation problem can be described as a black box problem, which is to determine the parity of the permutation based on the $n$ input and output states of black box. For example, in a set of six different possible permutations  (1,2,3),  (2,3,1),  (3,1,2),  (3,2,1),  (2,1,3) and  (1,3,2), the first three are positive cyclic or even permutations while the last three are negative cyclic or odd permutations. If we treat permutation operation as a function $ f (x) $ defined on the set $x\in  \left\lbrace 1,2,3\right\rbrace$, determination of parity requires the evaluation of $ f (x) $ for at least two different input values of $x$ with classical algorithm.

In this article, we consider four orthonormal states, i.e., $\vert 1\rangle= (1,0,0,0)^{T}$,  $\vert 2\rangle= (0,1,0,0)^{T}$, $ \vert 3\rangle= (0,0,1,0)^{T}$, and $\vert 4\rangle= (0,0,0,1)^{T}$, and the bijection is $\textit{f} :\lbrace1,2,3,4\rbrace\longrightarrow\lbrace1,2,3,4\rbrace $. For the input state  (1,2,3,4), there are eight different possible output states. These eight states and the corresponding transformations are divided into two categories as mentioned above. The positive cyclic permutations and the corresponding unitary transformations are:
\begin{eqnarray}
&&\ f_{1}=\left (\begin{array}{cccc} 1 & 2 & 3 & 4 \\ 1 & 2 & 3 & 4
\end{array} \right),  
\ \ f_{2}=\left (\begin{array}{cccc} 1 & 2 & 3 & 4 \\ 2 & 3 & 4 & 1
\end{array} \right),
\ f_{3}=\left (\begin{array}{cccc} 1 & 2 & 3 & 4 \\ 3 & 4 & 1 & 2
\end{array} \right),
\ \ f_{4}=\left (\begin{array}{cccc} 1 & 2 & 3 & 4 \\ 4 & 1 & 2 & 3
\end{array} \right), \notag\\
&&\ U_{1}=\left (\begin{array}{cccc} 1 & 0 & 0 & 0 \\ 0 & 1 & 0 & 0 \\ 0 & 0 & 1 & 0 \\ 0 & 0 & 0 & 1
\end{array} \right),
\ U_{2}=\left (\begin{array}{cccc} 0 & 1 & 0 & 0 \\ 0 & 0 & 1 & 0 \\ 0 & 0 & 0 & 1 \\ 1 & 0 & 0 & 0
\end{array} \right),
\ U_{3}=\left (\begin{array}{cccc} 0 & 0 & 1 & 0 \\ 0 & 0 & 0 & 1 \\ 1 & 0 & 0 & 0 \\ 0 & 1 & 0 & 0 
\end{array} \right),
\ U_{4}=\left (\begin{array}{cccc} 0 & 0 & 0 & 1 \\ 1 & 0 & 0 & 0 \\ 0 & 1 & 0 & 0 \\ 0 & 0 & 1 & 0
\end{array} \right),
\label{eq:1}
\end{eqnarray}
and the negative cyclic permutations and the corresponding unitary transformations are:
\begin{eqnarray}
&&\ f_{5}=\left (\begin{array}{cccc} 1 & 2 & 3 & 4 \\ 4 & 3 & 2 & 1
\end{array} \right),
\ \ f_{6}=\left (\begin{array}{cccc} 1 & 2 & 3 & 4 \\ 3 & 2 & 1 & 4
\end{array} \right),
\ f_{7}=\left (\begin{array}{cccc} 1 & 2 & 3 & 4 \\ 2 & 1 & 4 & 3
\end{array} \right),
\ \ f_{8}=\left (\begin{array}{cccc} 1 & 2 & 3 & 4 \\ 1 & 4 & 3 & 2
\end{array} \right), \notag\\
&&\ U_{5}=\left (\begin{array}{cccc} 0 & 0 & 0 & 1 \\ 0 & 0 & 1 & 0 \\ 0 & 1 & 0 & 0 \\ 1 & 0 & 0 & 0
\end{array} \right),
  \ U_{6}=\left (\begin{array}{cccc}  0 & 0 & 1 & 0 \\ 0 & 1 & 0 & 0 \\ 1 & 0 & 0 & 0 \\ 0 & 0 & 0 & 1
\end{array} \right),
  \ U_{7}=\left (\begin{array}{cccc}  0 & 1 & 0 & 0 \\ 1 & 0 & 0 & 0 \\ 0 & 0 & 0 & 1 \\ 0 & 0 & 1 & 0
\end{array} \right),
  \ U_{8}=\left (\begin{array}{cccc}  1 & 0 & 0 & 0 \\ 0 & 0 & 0 & 1 \\ 0 & 0 & 1 & 0 \\ 0 & 1 & 0 & 0
\end{array} \right).
\label{eq:1}
\end{eqnarray}
The black box can be considered as a special device which can realize the corresponding operation for a given permutation task. 

To determine the parity, a direct running of the operator on the eigen states of ququart will not work. Therefore at least twice runnings are needed to evaluate the box parity as in the classical case. For example, if we input a state $\vert 2\rangle$ and get the output state $\vert 4\rangle$, this progress corresponds to two possible permutation transformations $ f_{3} $ and $ f_{8} $. So to determine the parity of permutation, we need at least another running.
\begin{figure}[hbt]
\centering
\includegraphics[width=13cm]{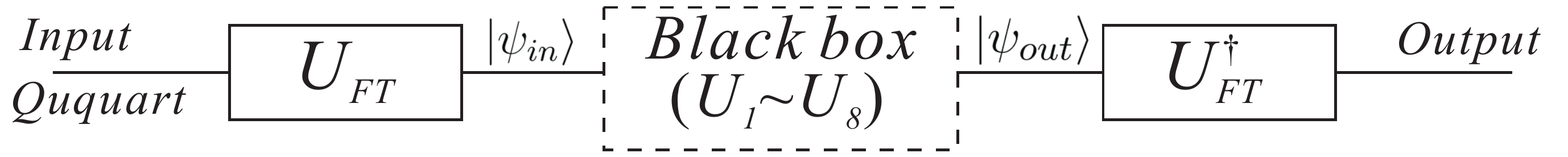}
\caption{\textbf{Quantum circuit to realize the permutation algorithm.}  $ U_{FT} $ denotes the Fourier transformation. The black box contains all the eight permutation operators in the algorithm and $U_{FT}^{\dagger}$ is the inverse Fourier transformation.}
\label{fig00}
\end{figure}

Fig.~\ref{fig00} gives the quantum circuit to realize the permutation algorithm. Considering the quantum algorithm of this task, we start from a superposition state:
\begin{equation}
\vert\psi_{in}\rangle=\vert\psi_{2}\rangle=\left( \vert 1\rangle+i\vert 2\rangle-\vert 3\rangle-i\vert 4\rangle\right)/2,
\end{equation}
here $\vert\psi_{j}\rangle=U_{FT}\vert j \rangle, (j=1,2,3,4)$ which is generated by quantum Fourier transform $U_{FT} $:
\begin{equation}
\ U_{FT}=\frac{1}{2}\left(\begin{array}{cccc}  1 & 1 & 1 & 1 \\ 1 & i & -1 & -i \\ 1 & -1 & 1 & -1 \\ 1 & -i & -1 & i
\end{array} \right).
\end{equation}
Then the input state get through the black box, and the output states   can be written as
\begin{equation}
\ \vert \psi_{out} \rangle_{f_{k}} \equiv U_{k} \vert \psi_{in} \rangle=\frac{\left( \vert f_{k}\left( 1\right) \rangle+i\vert f_{k}\left( 2\right) \rangle-\vert f_{k}\left( 3\right) \rangle-i\vert f_{k}\left( 4\right) \rangle\right)}{2},
\end{equation}
where $ k=1,2,\cdots ,8 $ label eight permutation transformations. Application of $ U_{k} $ on the input state $ \vert \psi_{in} \rangle $ gives $\vert\psi_{2}\rangle$ for even $ f_{k} $ and it gives $\vert\psi_{4}\rangle$ for odd $ f_{k} $. 
By employing the inverse Fourier transform $U_{FT}^{\dagger}$ and checking the final state, we will know the parity of permutation is even (odd) when we acquire the state $\vert 2\rangle$ $(\vert 4\rangle)$. Thus, the quantum algorithm allows us to determine the parity of a cyclic permutation with single evaluation of the permutation function instead of two.

Reviewing this algorithm, the Fourier transform $U_{FT}$ and inverse Fourier transform $U_{FT}^{\dagger}$ are not necessary if we can directly prepare the superposition state $ \vert \psi_2\rangle$ and discriminate the output state $\vert\psi_{2}\rangle$ and $\vert\psi_{4}\rangle$. Luckily, by employing the photon's polarization and spatial mode, we can easily realize the ququart $\vert\psi_{2}\rangle$ and distinguish $\vert\psi_{2}\rangle$ and $\vert\psi_{4}\rangle$ in our experiment. We should address that $\vert\psi_{4}\rangle$ is also a proper input state for the permutation discrimination, and in our experiment, we demonstrate the algorithm for both $\vert\psi_{2}\rangle$ and $\vert\psi_{4}\rangle$.
\\
\\
\noindent \textbf{Results}

\noindent In this context,
we utilize photon's  polarization and spatial mode to code the ququart \cite{Zhang2010}, and we carry out the whole eight different permutations in a black box which is composed by Dove prism (DP) and half wave plates (HWP). Combining the polarization and spatial mode as composite quantum state, the four states of ququart can be defined as 
\begin{equation}
\begin{array}{cc}   
\vert H,l \rangle \rightarrow \vert 1\rangle,\  
\vert H,r \rangle \rightarrow \vert 2\rangle,\ 
\vert V,l \rangle \rightarrow \vert 3\rangle,\ 
\vert V,r \rangle \rightarrow \vert 4\rangle,
\end{array}
\end{equation}
where $ H \ (V) $ represents horizontal (vertical) polarization and $  r \ (l) $ represents right (left) spatial mode. 
The input state $\vert \psi_{2} \rangle$ can be written in the form
\begin{equation}
\vert \psi_{2} \rangle=\vert H,l \rangle+i\vert H,r \rangle-\vert V,l \rangle-i\vert V,r \rangle.
\label{eq06}
\end{equation}
Then the black box will carry out the permutation transformation on $ \vert \psi_{2} \rangle $. For example, suppose the operation is $ f_{2} $, the output state after black box can be expressed as $\vert H,r \rangle+i\vert V,l \rangle-\vert V,r \rangle-i\vert H,l \rangle,$ which is equal to $-i\vert \psi_{2} \rangle. $
\begin{figure}[hbt]
\centering
\includegraphics[width=12cm]{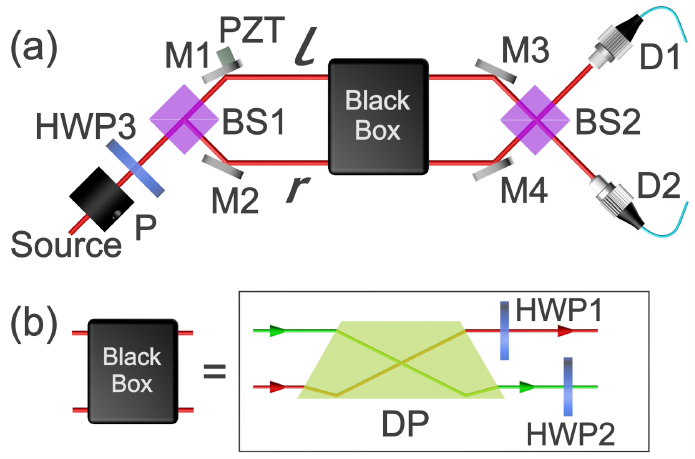}
\caption{\textbf{Experimental setup of permutation algorithm.} (a) The single photon source is achieved by deeply attenuating coherent light into single photon level. P denotes the polarizer for preparing horizontal polarization state. HWP3 are used for initial polarization state preparation.  Two beam splitters (BS) and four mirrors (M) are used to set up a Mach-Zehnder interferometer. The piezo transmitter (PZT) actuator is used to modulate the phase $ \varphi $ between $ l $ and $  r $ paths. The black box is used to realize eight permutation transformations. Two detectors (D1 and D2) are single-photon counting module (SPCM-AQRH-14-FC) used to record the count of photons. (b) The black box consists of a DP at $0^{\circ}$ and two HWPs at $45^{\circ}$. Different permutation transformation can be achieve by different combination of DP and HWPs.}
\label{fig01}
\end{figure}

The sketch of the experimental setup is shown in Fig.~\ref{fig01}(a). The source is achieved by deeply attenuating coherent light into single photon level (coherent parameter $ \alpha\approx 0.1 $). Polarizer (P) and HWP3 are placed to prepare the initial input polarization state and beam splitter (BS1) is to prepare spatial state. The black box consists of a DP and two HWPs as shown in Fig.~\ref{fig01}(b), the dove prism is located at an angle $0^{\circ}$ relative to the horizontal plane. Photons with different spatial modes \textit{r} and \textit{l} pass through DP will swap their spatial modes, and the polarization will be altered by HWP at at $45^{\circ}$ succinctly. To realize the permutation transformations with black box, what we need is adding or removing the DP and HWPs. 

In this scheme, we prepare the initial polarization state of photon in a superposition state $\left( \vert H \rangle - \vert V \rangle \right)/\sqrt{2}$, then we transform the spatial mode into  $\left( \vert l \rangle + \vert r \rangle\right) /\sqrt{2}$ by BS1. The relative phase $\varphi$ between two spatial modes can be adjusted by a PZT mounted on the first reflecting mirror (M1). Thus the input state for the black box can be written as
\begin{equation}
\begin{aligned}
\vert \psi_{in} \rangle&=\left ( \vert H \rangle-\vert V \rangle\right)\left ( \vert l \rangle+e^{i\varphi}\vert r \rangle\right)/2 \\ 
&=\left( \vert H,l \rangle+e^{i\varphi}\vert H,r \rangle-\vert V,l \rangle-e^{i\varphi}\vert V,r \rangle\right)/2.
\end{aligned}
\end{equation}
Especially, when $\varphi=\pi/2$, the initial state $\vert \psi_{in} \rangle$ turns to be $\vert \psi_{2} \rangle$ which is expressed in Eq.~(\ref{eq06}); when $\varphi=3\pi/2$, the initial state $\vert \psi_{in} \rangle$ turns to be $\vert \psi_{4} \rangle$. Then the state is injected into the black box to carry out the permutation transformations. The states of $\vert \psi_{2} \rangle$ and $\vert \psi_{4} \rangle$ 
in our definition can be discriminated directly by the phase of spatial mode. So different spatial states interfere at the second beam splitter (BS2), and the parity of permutation can be determined by checking which single photon detector (D1 or D2) clicks.
\begin{table}[hbt]
\centering
\caption{ Experimental implementation of eight different permutations}
\label{table:1}
\begin{threeparttable}
\begin{tabular}{|c|c|c|c|c|c|c|c|c|c|}
\hline
\multicolumn{2}{|c|}{\large State}  & \ \ \  H,l \ \ \ & \ \ \ H,r \ \ \ & \ \ \ V,l \ \ \ & \ \ \ V,r \ \ \ &  \multicolumn{3}{c|}{\large Implementation} & \multirow{2}{0.8in}{\ \ \large Parity}  \\ \cline{1-9} 
\multicolumn{2}{|c|}{\large Code}  &  1  &  2  &  3  &  4 & \ \ \ \ \ \  DP \ \ \ \ \ \  & \ \ \ \ \ \  HWP1 \ \ \ \ \ & \ \ HWP2 \ \ & \\ \hline  
\multirow{8}{0.5cm}{\rotatebox{270}{\large Bijection}}  & \ \ \ \ $f_{1}$\ \ \ \  & H,l & H,r & V,l & V,r & No & No & No & even\\ \cline{2-10} 
& $f_{2}$  &  H,r  &  V,l  &  V,r  &  H,l  & Yes & Yes & No  & even\\ \cline{2-10}
& $f_{3}$  &  V,l  &  V,r  &  H,l  &  H,r  & No & Yes & Yes  & even \\ \cline{2-10}  
& $f_{4}$  &  V,r  &  H,l  &  H,r  &  V,l  & Yes & No & Yes & even \\ \cline{2-10} 
& $f_{5}$  &  V,r  &  V,l  &  H,r  &  H,l  & Yes & Yes & Yes & odd \\ \cline{2-10} 
& $f_{6}$  &  V,l  &  H,r  &  H,l  &  V,r  & No & Yes & No & odd \\ \cline{2-10}
& $f_{7}$  &  H,r  &  H,l  &  V,r  &  V,l  & Yes & No & No  & odd \\ \cline{2-10}
& $f_{8}$  &  H,l  &  V,r  &  V,l  &  H,r  & No & No & Yes & odd \\ \hline 
\end{tabular}
\end{threeparttable}%
\end{table} 

\noindent \textbf{Implementation of positive cyclic permutation.}
For the identity operation $f_{1}$, we remove the DP and HWPs, the input states pass through the setup remain unchanged and we achieve the $f_{1}$ operation. For the $f_{2}$ operation,  DP and HWP1 are required in the setup. When the photons in different paths pass through the DP, they will exchange their paths ($l\rightarrow r$ and $r\rightarrow l$), and the state $\vert \psi_{in} \rangle$ changes into $\vert H,r \rangle+e^{i\varphi}\vert H,l \rangle-\vert V,r \rangle-e^{i\varphi}\vert V,l \rangle $. After that, HWP1 at an angle $45^{\circ}$ is located at the \textit{l} path, it will alter horizontal polarization to vertical polarization and vice versa. So far we get the final state $\vert H,r \rangle+e^{i\varphi}\vert V,l \rangle-\vert V,r \rangle-e^{i\varphi}\vert H,l \rangle $ which is equal to $ -i\vert \psi_{2}\rangle $ when $ \varphi=\pi/2$ and achieve the mapping operation $f_{2}$. The $f_{3}$ operation can be carried out similarly by removing the DP and putting HWP1 and HWP2. Analogous to the action above, the last positive cyclic permutation $f_{4}$ can be accomplished by inserting  DP and HWP2. When photons pass through the device for the positive cyclic permutation, a simple calculation shows the output state $\vert \psi_{out} \rangle=U_{k}\vert \psi_{in} \rangle \  (k=1,2,3,4)$ as follows:
\begin{small}
\begin{equation}
\begin{aligned}
&\vert \psi_{out} \rangle_{f_{1}}=\left ( \dfrac{\vert H \rangle-\vert V \rangle}{\sqrt{2}}\right)\left (\dfrac{\vert l \rangle+e^{i\varphi}\vert r \rangle}{\sqrt{2}} \right); \ \
\vert \psi_{out} \rangle_{f_{2}}=-e^{i\varphi}\left ( \dfrac{\vert H \rangle-\vert V \rangle}{\sqrt{2}}\right)\left (\dfrac{\vert l \rangle+e^{i\left (\pi-\varphi \right)}\vert r \rangle}{\sqrt{2}} \right) \\
&\vert \psi_{out} \rangle_{f_{3}}=-\left ( \dfrac{\vert H \rangle-\vert V \rangle}{\sqrt{2}}\right)\left (\dfrac{\vert l \rangle+e^{i\varphi}\vert r \rangle}{\sqrt{2}} \right);\ \
\vert \psi_{out} \rangle_{f_{4}}=e^{i\varphi}\left ( \dfrac{\vert H \rangle-\vert V \rangle}{\sqrt{2}}\right)\left (\dfrac{\vert l \rangle+e^{i\left (\pi-\varphi \right)}\vert r \rangle}{\sqrt{2}} \right).
\end{aligned} 
\label{eq08}
\end{equation}
\end{small}

\noindent\textbf{Implementation of negative cyclic permutation.}  Firstly we  place both DP and HWPs into the optical route to exchange the spatial mode and the polarization.  This will be the $f_{5}$ gate. If we remove the DP and only place HWP1, when the photons pass this setup, it will go through an $f_{6}$ transformation. For the purpose of carrying out $f_{7}$ operation, it is easy to obtain that only a DP is needed. The last operation can be achieved by only employing HWP2 to change the polarization mode on the right route.
Similarly, we arrive the final output states:
\begin{small}
\begin{equation}
\begin{aligned}
&\vert \psi_{out} \rangle_{f_{5}}=-e^{i\varphi}\left ( \dfrac{\vert H \rangle-\vert V \rangle}{\sqrt{2}}\right)\left (\dfrac{\vert l \rangle-e^{i\left(\pi- \varphi\right) }\vert r \rangle}{\sqrt{2}} \right); \ \
\vert \psi_{out} \rangle_{f_{6}}=-\left ( \dfrac{\vert H \rangle-\vert V \rangle}{\sqrt{2}}\right)\left (\dfrac{\vert l \rangle-e^{i\varphi}\vert r \rangle}{\sqrt{2}} \right) \\
&\vert \psi_{out} \rangle_{f_{7}}=e^{i\varphi}\left ( \dfrac{\vert H \rangle-\vert V \rangle}{\sqrt{2}}\right)\left (\dfrac{\vert l \rangle-e^{i\left(\pi- \varphi\right) }\vert r \rangle}{\sqrt{2}} \right); \ \
 \vert \psi_{out} \rangle_{f_{8}}=\left ( \dfrac{\vert H \rangle-\vert V \rangle}{\sqrt{2}}\right)\left (\dfrac{\vert l \rangle-e^{i\varphi}\vert r \rangle}{\sqrt{2}} \right).
\end{aligned} 
\label{eq09}
\end{equation} 
\end{small}

From the above discussion, we carry out all the eight essential transformations for the parity determining algorithm. All the transformations and corresponding implementation approaches are summarized in Table \ref{table:1}.
In particular, when $\varphi=\dfrac{\pi}{2}$, the output states in Eqs.~(\ref{eq08}) and~(\ref{eq09}) are equal to $ \vert \psi_{2}\rangle $ and  $\vert \psi_{4}\rangle $, respectively. The Eqs.~(\ref{eq08}) and~(\ref{eq09}) clearly show that the polarization of output states are same while the relative phase between two spatial modes are different for two types of parity. This means that we can accomplish the identify progress of permutation through the spatial bits without the inverse Fourier transform. After the spatial modes interfered on BS2, we can determine the parity is odd (even) when detector D1 (D2) clicks. With similar analysis, when the input state is $ \vert \psi_{4}\rangle $ (relative phase $\varphi=\dfrac{3\pi}{2}$), we can determine the parity is odd (even) when detector D2 (D1) clicks.

\begin{figure}[hbt]
  \centering 
  \subfigure{
    \label{fig:subfig:a} 
    \includegraphics[width=15cm]{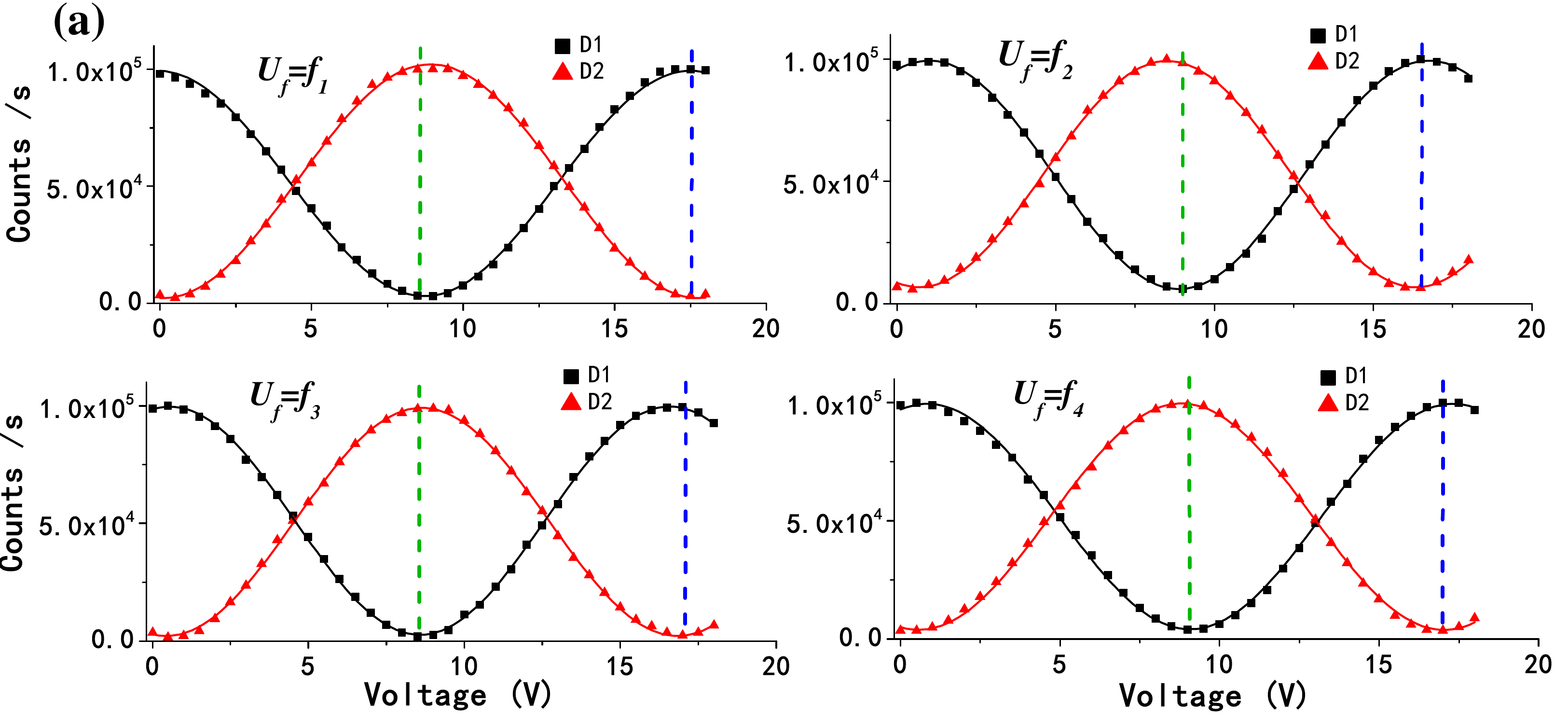}}
  \hspace{0.01in}
  \subfigure{
    \label{fig:subfig:b} 
    \includegraphics[width=15cm]{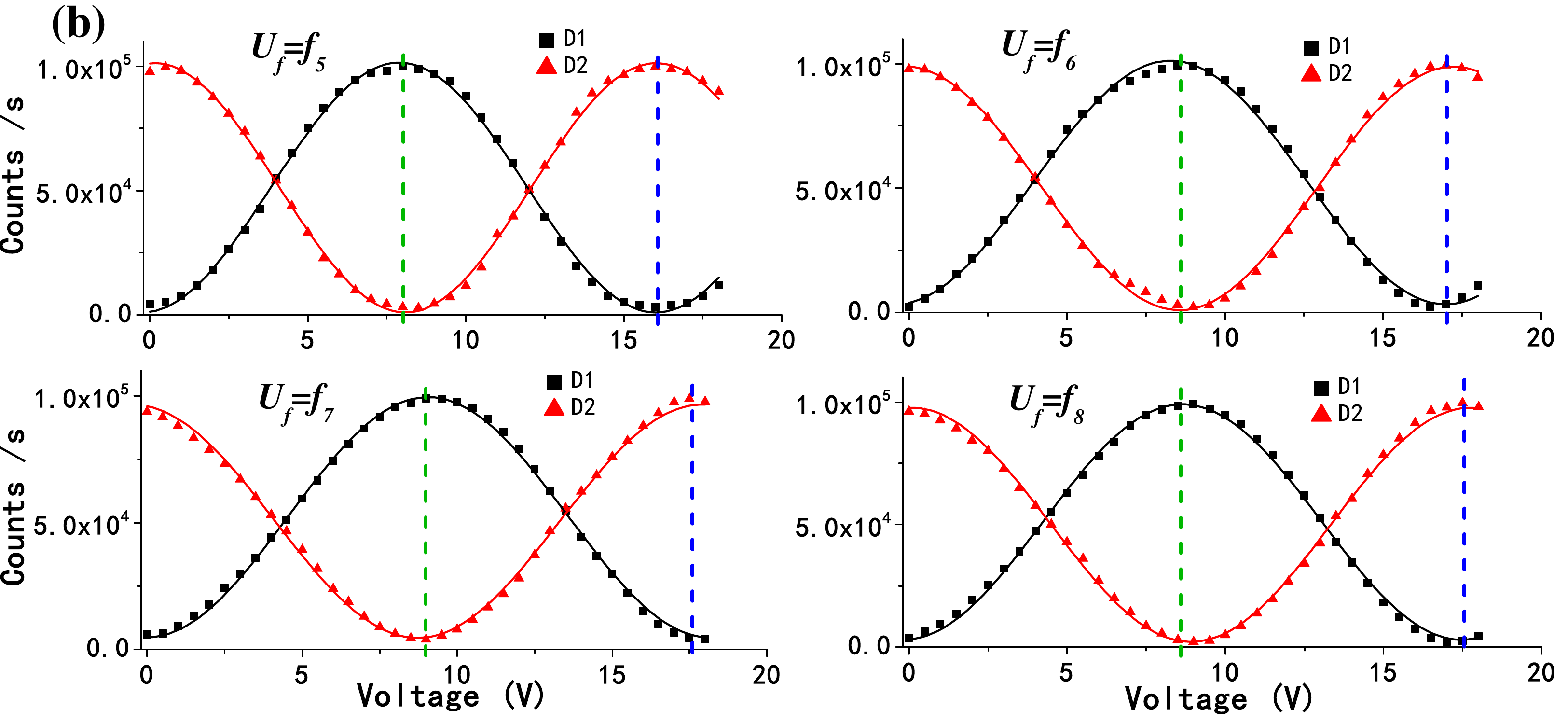}}
  \caption{\textbf{Experimental data of the parity determining algorithm.}  (a) positive cyclic permutation operation, (b) negative cyclic permutation operation. Black square dots show the photon counts of D1 , and red triangle points show the photon counts of D2. Fitting lines are also shown. Green (blue) dashed vertical lines are used to mark the proper points  (phases) of the initial states $ \vert \psi_{2}\rangle $ ($ \vert \psi_{4}\rangle $), which can be used to perfectly discriminate the two kinds of functions.}
  \label{fig03} 
\end{figure}

We record the photon counts of the two detectors  D1 and D2 with 0.5 V a step of the PZT voltage synchronously. 
Our experimental results are shown in Fig.~\ref{fig03}. Figure~\ref{fig:subfig:a} shows the results of positive cyclic permutation transformations from $f_{1}$ to $f_{4}$, and Fig.~\ref{fig:subfig:b} gives the results of the remaining four odd operations. The black square spots represent the counts of D1, and the red triangular spots corresponding to the counts of D2. As we discussed above, when the relative phase $\varphi$ is equal to $\left (2N+1/2 \right) \pi$, where N is an integer, only detector D1 clicks for the odd parity and D2 clicks for the even parity. These special points are pointed out by the green dashed line, and from these points we can get the parity information of the permutation evidently. Analogously, when the relative phase $\varphi$ is equal to $(2N + 3/2) \pi$, we still can determine the parity by the blue dashed line labelled in Fig. 3.
\\
\\
\noindent \textbf{Discussion}

\noindent  We define $\eta=\vert C_{D1}-C_{D2} \vert / \vert C_{D1}+C_{D2} \vert$ as the contrast ratio to evaluate the accuracy of our experiment, where $ C_{D1} $ and $ C_{D2} $  denote the photon counts of D1 and D2, respectively. Theoretically, the contrast ratio $ \eta $ is equal to 1. In our experiment, The contrast ratio is up to  $96 \pm2 \% $  for all eight cases in Fig.~\ref{fig03}. The error and the small shift of the dashed line for different permutations mainly comes from the imperfection of optical devices and moving or adding DP and HWPs.

In conclusion, we briefly introduce the quantum permutation algorithm and put forward a scheme to implement this algorithm by employing linear optical element. The main point to achieve the parity determining algorithm is the bijection operator and recognize the parity through the output states from the two detectors. 
 By using composite quantum bit to realize ququart, our experiment is greatly simplified both in state preparation and state discrimination, where Fourier transformations $ U_{FT} $ and  $ U_{FT}^\dagger $  are saved. Although this algorithm only provide a two to one speed-up towards classical case, it shows the  power of quantum parallelism and quantum computation validly, and expand our thinking for more efficient algorithms.  We noticed that this scheme has also been realized in spin-$\frac{3}{2}$ NMR quadrupolar system with four energy levels \cite{Dogra2014,Silva2014}.
\\
\\
\noindent \textbf{Method} 

\noindent Figure~\ref{fig01} displays the setup needed in the experimental progress. For the specific operation such as \textit{$f_{2}$}, we only need one HWP at $45^{\circ}$ in the left route, however, this HWP gives an additional phase because of the intrinsic thickness. With the purpose of eliminating this extra effects, we put another HWP at $0^\circ$ in the right path to compensate phase without changing the polarization. 
We also need to calibrate the voltage of PZT to confirm the initial state is proper prepared before the experiment proper starts.

\noindent
\\
\noindent \textbf{Acknowledgements}

\noindent This work is supported by the Fundamental Research Funds for the
Central Universities and the National Natural Science Foundation of China ( Grant Nos. 11374008, 11374238 and 11374239).
\\
\\
\noindent \textbf{Author contributions}

\noindent P.Z. conceived and designed the experiment. F.W. and Y.W. performed the experiment and collected the data. F.W., R.L. and D.C. carried out theoretical calculations.  F.W., Y.W., R.L., P.Z. H.G. and F.L. contributed to writing the manuscript.
\\
\\


\end{document}